\documentclass[a4paper,10pt]{article}
\usepackage[utf8]{inputenc}
\usepackage{graphicx}
\usepackage{hyperref}
\title{Modified Theory of Gravity and Clustering of Multi-Component System of Galaxies}
\author{{Mir Hameeda$^{a}$\thanks{E-mail: hme123eda@gmail.com}~,~ Behnam Pourhassan$^{b}$\thanks{Email: b.pourhassan@du.ac.ir}~,~
Mir Faizal$^{c,d}$\thanks{Email: mirfaizalmir@googlemail.com},}\\
{C. P. Masroor$^e$\thanks{E-mail: masroorcp@gmail.com}~,~ Rizwan Ul Haq Ansari$^f$\thanks{E-mail: rizwan@manuu.ac.in}~,~ P. K. Suresh$^e$\thanks{E-mail: pkssp@uohyd.ernet.in}}\\
$^{a}${\small {Department of Physics, S.P. Collage,  Srinagar, Kashmir, 190001, India}}\\
$^{a}${\small { Visiting Associate, IUCCA,  Pune,  41100, India}}\\
$^{b}${\small {\em School of Physics, Damghan University, Damghan, 3671641167, Iran}}\\
$^{c}${\small {\em Deptartment of Physics and Astronomy, University of Lethbridge,}}\\
{\small {\em Lethbridge, AB T1K 3M4, Canada}}\\
$^{d}${\small {\em Irving K. Barber School of Arts and Sciences,  University of British Columbia - Okanagan,}}\\
{\small {\em Kelowna, BC V1V 1V7, Canada}}\\
$^{e}${\small {\em School of Physics,  University of Hyderabad}}\\
{\small {\em Hyderabad, 500 046, India}}\\
$^{f}${\small {\em Department of Physics, Maulana Azad National Urdu University}}\\
{\small {\em Hyderabad, 500 032, India.}}}
\begin{document}

\maketitle
\begin{abstract}
In this paper, we analyze the clustering of  galaxies using a modified theory of gravity,
in which the field content of general relativity has been be increased. This increasing in the field content of general relativity
 changes the large distance
behavior of the theory, and  in weak field approximation,  it will also modify the large distance behavior of Newtonian potential.
So, we will analyzing the clustering of multi-component system of  galaxies interacting through this   modified
Newtonian potential. We will obtain the partition function for this multi-component system,
and study the thermodynamics of this system. So, we will analyze the effects of the large distance modification to the Newtonian
potential on Helmholtz free energy, internal energy, entropy,
pressure and chemical potential of this system.   We obtain also the modified  distribution function and the
modified  clustering parameter for this system, and
hence observe the effect of large distance modification of Newtonian potential  on clustering of galaxies.
\end{abstract}

\section{Introduction}
Observations made on the dynamics of galaxies indicate a discrepancy between the observed mass of galaxy from
dynamics of galaxies and the mass inferred from
the existence of luminous matter \cite{rubin1, rubin2}. It appears that a large part of the
mass of the galaxies and thus universe as a  whole universe is not visible, this non-luminous missing mass of the universe is known as  dark
matter \cite{dark1, dark2}. Several models have been proposed for the dark matter, such as
 axion \cite{axion}, black holes \cite{black hole}, neutrino \cite{neutrino} and gravitino \cite{gravitino}.
 However, none of these dark matter models has been verified, and this has led to the development of alternative approaches to
 explain this discrepancy between the observed and measured rotation of galaxies. These approaches are motivated by a large
 distance modification of dynamics, such that this modified dynamics can resolve this discrepancy.
 In fact, it has been demonstrated that by modifying the Newtonian dynamics at galactic scales, it is possible to resolve this
 discrepancy, and this large distance correction to the Newtonian  dynamics  is called
 Modified Newtonian Dynamics (MOND) \cite{milgrom}. Even though the MOND explains the dynamics at galactic scale, it
 does not correct describe the dynamics at intra-galactic scale, and hence it cannot be used to analyze the clustering
 of galaxies \cite{1, 2, 4}.

 It has also been argued that it is possible to have other modifications to gravity, such as
 modified theory of gravity  (MOG), which do not have this problem, and  can explain the clustering of galaxies \cite{5, 6}.
 In MOG, the field content of general relativity are increased to include   scalar,   and  vector fields, apart from the tensor field
 \cite{moffat06}.  The dynamics of a test particle in MOG are modified by the inclusion of these additional fields.
 This is because the  coupling of the metric to   both the scalar   and the vector fields,
 modifies the usual  solution to the field equations for a point mass~\cite{moffat09}. In fact, the   rotation curves of galaxies
in MOG have also been analyzed using a static spherically symmetric
point mass solution derived from the field equations \cite{brown2006, brown2009}. The same procedure has been applied to the dynamics of globular
clusters \cite{moffattoth}, clusters of galaxies \cite{brownMoffat2006},  and the bullet cluster \cite{brownMoffat2007}.
It has been observed that MOG can explain the dynamics at intra-galactic scales, and hence it can be used to analyze clustering
of galaxies. In the weak field approximation, MOG produces the   Newtonian potential with
    and a large distance correction to the Newtonian potential.
The  weak field approximation of the MOG has been used for  analyzing such systems  \cite{w, w1, mmf},
   so in this paper,  we will also use the weak field approximation to MOG.

   As the intra-galactic distances are much larger than the diameter of individual galaxies, we can approximate the
individual galaxies   as point particle  \cite{sas85}.
So, such  a  system of galaxies interacting through a potential can be studied using
standard techniques of
statistical mechanics. In fact, such a system of galaxies interacting thought a Newtonian potential has already been studied
using such techniques from
statistical mechanics \cite{ahm10, sas84, ahm06}.
It has also been observed that for this system  the
gravitational clustering can evolve through a sequence of quasi-equilibrium state \cite{ahm10, ahm12}.
Thus, the  cosmological many body partition function has been obtained by using  an
ensemble of co-moving cells containing   galaxies interacting through the usual Newtonian potential \cite{ahm02}.
It has also been observed that if the cells is smaller than the  correlation length, then each member of this ensemble is correlated
gravitationally with other cells. So, for such cells, the correlations within a cell is greater than correlations among cells,
so that extensivity is a good approximation to study such a system \cite{ahm02, sas00}.
The techniques of statistical mechanics can also  used to analyze the clustering of different types of galaxies. In fact,
  the clustering of different kind of galaxies has been studied using a multi-component system \cite{ahm14, mu12, mu14}.
The galaxies in this multi-component system were  again assumed to interact through a Newtonian potential.

The clustering has also been studied using the large distance modification of Newtonian potential. It is possible to obtain large distance
correction to the Newtonian potential in brane world models \cite{mnpld},  and the clustering of galaxies has been studied using such a large
distance correction to Newtonian potential \cite{Hameeda:2015gra}. The effects of    of cosmological constant on clustering of galaxies
has   been analyzed, and the thermodynamics for such a system of galaxies has been studied
\cite{Hameeda:2016xxs}. The   cluster of galaxies under the effect of dynamical dark energy has also been
studied, and the  gravitational partition function for this system has been constructed \cite{Pourhassan:2017zjs}.
This gravitational partition function has been used to analyze the thermodynamics of this system. As the dark energy is dynamical
in this model, the  time evolution of the clustering parameter  is studied using the  time dependence of this dynamical dark energy.
So, it is both interesting and important to analyze the large distance modification to the clustering of galaxies using standard
techniques of statistical mechanics. As MOG produces an phenomenologically important large distance modification of Newtonian
potential, in this paper, we will use this MOG modified Newtonian potential to analyze the clustering of galaxies.

\section{Modified Newtonian Potential}
In this section, we obtain the MOG modified potential for a system of  different types of galaxies. So, first we will review the
weak field approximation of MOG \cite{w, w1, mmf}, and then analyze the modification to that potential from the
softening parameter which is used for studding the clustering of galaxies  \cite{ahm10, ahm02, ahm06}. Apart  from tensor field, MOG
consists of massive vector field $\phi_\mu$ and three scalar fields, Newton's constant $G$, a vector field coupling
constant $\omega$. The mass of the vector filed $\mu$ acts as scalar fields, and so the mass of the scalar field is a dynamical
function in  space-time.
This theory also contains the self interacting potentials for various field, which can be denoted by
$V_\phi(\phi_\mu\phi^\mu)$, $V_G(G)$, $V(\omega)$ and $V_\mu(\mu)$.
Now the action for MOG can be written as \cite{moffat06, mmf},
\begin{equation}\label{action1}
S=S_G+S_\phi+S_S+S_M,
\end{equation}
where $S_G$ is the original  Einstein gravity action, $S_\phi$ is the action of a massive vector field $\phi$, $S_S$ is the action of
scalar fields, and $S_M$ is the matter action, which can be considered as pressureless dust. These actions can be expressed as
\begin{eqnarray}
 S_G&=&-\frac{1}{16\pi}\int\frac{1}{G}\left({\it R}+2\Lambda\right)\sqrt{-g}~d^4x,\nonumber \\
 S_\phi&=&-\frac{1}{4\pi}\int\omega\Big[\frac{1}{4}{\bf\it B^{\mu\nu}B_{\mu\nu}}-\frac{1}{2}\mu^2\phi_\mu\phi^\mu+ V_\phi(\phi_\mu\phi^\mu)\Big]\sqrt{-g}~d^4x,\nonumber \\
S_S&=&-\int\frac{1}{G}\Big[\frac{1}{2}g^{\alpha\beta}\biggl(\frac{\nabla_\alpha G\nabla_\beta G}{G^2}
+\frac{\nabla_\alpha\mu\nabla_\beta\mu}{\mu^2}\biggr)+\frac{V_G(G)}{G^2}+\frac{V_\mu(\mu)}{\mu^2}\Big]\sqrt{-g}~d^4x,
 \nonumber \\
 S_M &=& \int(- \rho \sqrt{u^\mu u_\mu} - \omega
Q_5u^\mu\phi_\mu)\sqrt{-g} dx^4,
\end{eqnarray}
where $B_{\mu \nu}= \partial_\mu \phi_\nu - \partial_\nu \phi_\mu$, and $\nabla_\nu$ is covariant derivative with
respect to metric $g_{\mu \nu}$. In matter action $\rho$ is matter density, and $Q_5$ is the source of the  fifth force,
and it is related to matter density as $Q_5 = \kappa \rho$, where $\kappa$ is a constant.

To make weak field approximation,  we have considered fields as background plus perturbation.
The indices $(0)$ and $(1)$ are background and perturbation respectively. Since, there is no gravitational
source for vector field $\phi_{\mu(0)}=0$, and $\phi_{\mu(1)} \equiv \phi_{\mu}$, the equation of motion for $G$ is given by
\begin{equation}
\nabla^\mu \nabla_\mu G_{(1)} = -\frac{G_{(0)}}{16\pi} R_{(1)},
\end{equation}
where $R_{(1)}$ is the perturbation of the Ricci scalar in the forth order.
Similarly, we get equation of motion varying the tensor component we get
\begin{equation}\label{ein1}
R_{\mu\nu(1)} - \frac{1}{2} R_{(1)} \eta_{\mu\nu} =  - 8\pi G_0 T_{\mu\nu(1)}^{(M)} - 8\pi G_0 T_{\mu\nu(1)}^{(\phi)}
\end{equation}
where we have considered only first order terms and $T_{\mu\nu(1)}^{(\phi)}$ is the energy momentum tensor of the vector field
\begin{eqnarray}\label{tphi}
T_{\mu\nu}^{(\phi)} &=& \frac{\omega}{4\pi}(B_\mu{}^{\alpha}
B_{\nu\alpha} - \frac{1}{4}
g_{\mu\nu}B^{\alpha\beta}B_{\alpha\beta})\nonumber \\
&-&\frac{\mu^2\omega}{4\pi}(\phi_\mu\phi_\nu -
\frac{1}{2}\phi_\alpha\phi^\alpha g_{\mu\nu}).
\end{eqnarray}
Here $T_{\mu\nu(1)}^{(M)}$ is the tensor for matter. Now considering $T_{\mu\nu(1)}^{(\phi)} \ll T_{\mu\nu(1)}^{(M)}$, we get
\begin{equation}
R_{(1)} =  8\pi G_0 T^\mu{}_{\mu(1)}^{(M)}.
\label{perturbR}
\end{equation}
Since we are considering pressureless matter, the energy momentum tensor $T^\mu{}_{\mu(1)}^{(M)} = \rho$.
For the scalar field $G$, we have
\begin{equation}
\nabla^2\biggl(\frac{G_{(1)}}{G_0}\biggr) =\frac{1}{2}G_0\rho.
\end{equation}
Here,
$G_{(1)}/G_0$ is of the order of the gravitational potential, and is of the order
$(v/c)^2$, where $v$ is the internal velocity of the system. So, for  clusters of
galaxies, the deviation from the constant $G_0$ is of the order of
$G_1/G_0\simeq 10^{-7}-10^{-5}$.

For $(0,0)$ component of first order perturbation of Ricci tensor can be written as
\begin{eqnarray}
R_{00(1)} =\frac{1}{2}\nabla^2 h_{00},
 \end{eqnarray}
and so, the equation of motion is
\begin{equation}
\frac{1}{2}\nabla^2( h_{00}) = -4\pi G_0 \rho.
\label{effpois}
\end{equation}

The equation of motion for massive vector field is
\begin{equation}
\nabla_\nu B^{\mu\nu} - \mu^2 \phi^\mu = -\frac{4\pi}{\omega} J^\mu.
\end{equation}
Assuming conservation of the vector matter current, $\nabla_\mu J^\mu=0$, makes it possible to impose the gauge condition,
$\phi^\mu{}_{,\mu} = 0$ in the weak field approximation. Thus, for the static case, we can write
\begin{equation}
\nabla^2\phi^0 -\mu^2\phi^0 = -\frac{4\pi}{\omega}J^0,
\end{equation}
and this has the solution,
\begin{equation}\label{phi0}
\phi^0(r ) = \frac{1}{\omega}\int\frac{e^{-\mu|{\bf r }-{\bf r '}|}}{|{\bf r }-{\bf r '}|}J^0({\bf r '})d^3{\bf r '}.
\end{equation}
The field equation for an effective potential in the weak field
approximation, can be obtained  as follows,
\begin{equation}\label{p2}
\nabla\cdot {\bf a}  - \frac12\nabla^2 h_{00} = - \omega\kappa
\nabla^2\phi^0,
\end{equation}
where ${\bf a}$ is acceleration of the test particle. An effective potential for the
test particle can be defined and given by, ${\bf a} = -\nabla\Phi_{eff}$, and relate it to    matter distribution as follows,
\begin{equation}
\nabla\cdot(\nabla\Phi_{eff} - \kappa\omega\nabla\phi^0) = 4\pi G_0\rho.
\end{equation}
The solution to the above Poisson equation, $\Phi_N$  is given by
\begin{equation}\label{poteffect}
\Phi_{N} = \Phi_{eff} - \kappa\omega\phi^{0}.
\end{equation}
So, the effective potential can be written as
\begin{equation}
\Phi_{eff}({\bf r }) = - \int\frac{G_0 \rho({\bf r '})}{|{\bf r }-{\bf r '}|}d^3{\bf r '} +\kappa^2\int
\frac{{\rm e}^{-\mu|{\bf r }-{\bf r '}|}}{|{\bf r }-{\bf r '}|}\rho({\bf r '})d^3{\bf x'}. \label{potential}
\end{equation}
From the above relation it is clear  that there is a  repulsive Yukawa force term, in addition to attractive
gravitational force. Using the Dirac-delta function $\rho({\bf r '})= M \delta^3({\bf r '})$, for a point massive particle the effective
potential becomes,
\begin{equation}
\Phi_{eff}(r ) = - \frac{G_0 M}{r } + \kappa^2\frac{M {\rm e}^{-\mu r }}{r },
\label{phieffpointmass}
\end{equation}
where $r =\vert{\bf r }\vert$. Expanding the exponential term for distances compared to $\mu^{-1}$, the effective potential becomes
\begin{equation}
\Phi_{eff}(r ) = - \frac{(G_0 - \kappa^2)M}{r } - \mu\kappa^2 M.
\end{equation}
The first term is the Newtonian gravitational contribution, $G_0 - \kappa^2 =G_N$. As at   large distances,    ($\mu r \rightarrow
\infty$), we just have the first term of this equation, so  $G_0$  can be identified with
the effective gravitational constant at
infinity $G_{\infty}$. The effective potential for an extended distribution of matter in
MOG, can be written as   \cite{w, w1, mmf}
\begin{equation}
\Phi_{eff}({\bf r }) = - G_\infty \left[\int\frac{\rho({\bf r '})}{|{\bf r }-{\bf r '}|}\biggl(1
-\frac{G_\infty - G_N}{G_\infty}{\rm e}^{-\mu|{\bf r }-{\bf r '}|}\biggr)d^3{\bf r '} \right].
\end{equation}
So, we can define   $\alpha = (G_\infty - G_N)/G_N$, and write the effective potential as
\begin{equation}
\Phi_{eff}(\vec r ) = - G_N \left[\int\frac{\rho(\vec r ')}{|\vec r -\vec r '|}(1+\alpha
-\alpha e^{-\mu|\vec r -\vec r '|})d^3r ' \right].
\end{equation}
where $\alpha$ and $\mu$ can be   treated as
constant parameters, in the weak field approximation  \cite{w, w1, mmf}. Now for a system of galaxies,
the MOG modified Newtonian potential  between two galaxies,
can be written as
\cite{moffat06},
\begin{eqnarray}
\phi_{i,j}=-\frac{Gm^2}{r_{ij}}\biggl(1+\alpha-\alpha e^{-\mu r}\biggr),
\end{eqnarray}
It may be noted that, for the point masses,  the partition function of galaxies interacting through usual Newtonian potential
 diverges at $r_{ij}=0$.
This divergence occurs due to the assumption that the galaxies are point like objects.
  However, this  divergence can be    removed by taking the extended nature of galaxies into account by introducing
a softening parameter which takes care of the finite size of each galaxy \cite{ahm10, ahm02, ahm06}.
Thus by incorporating the softening parameter
the MOG modified interaction potential energy between galaxies can be represented as,
\begin{eqnarray}
\phi_{i,j}=-\frac{Gm^2}{(r_{ij}^2+\epsilon^2)^{1/2}}\biggl(1+\alpha-\alpha e^{-\mu{r_{ij}}}\biggr).
\end{eqnarray}
This is the potential we will use for analyzing the clustering of galaxies,  in this paper.
\section{ Gravitational Partition Function}
In this section, we will first approximate galaxies as point particles, as the intra-galactic distances are much larger than
the galactic scales. However, we will analyze this using a   multi-component system as
  we can have different types of galaxies, with different parameters. The multi-component system has already been used to analyze
  different types  of galaxies  interacting though a Newtonian potential \cite{ahm14, mu12, mu14}, and here
  we will analyze such a system of galaxies interacting through a MOG modified Newtonian potential. To
  analyze such a multi-component system, we will first explicitly
  analyze  a three component system, which will  consists of $N_{1}$ galaxies of mass $m_{1}$, $N_2$
galaxies of mass $m_2$ and $N_3$ galaxies of mass $m_3$. The partition function
for such a system can be expressed as
\begin{eqnarray}
Z_{N}(T,V)&=& \frac{1}{\Lambda^{3N}N!}\int d^{3N_1}p_{i}d^{3N_1}rd^{3N_2}p_{j}d^{3N_2}r^\prime d^{3N_3}p_{k}d^{3N_3}r^{\prime\prime} \nonumber \\ &&
\times  exp\biggl(-\biggl[\sum_{i=1}^{N}\frac{p_{i}^2}{2m_1}+\sum_{j=1}^{N}\frac{p_{j}^2}{2m_2}+\sum_{k=1}^{N}\frac{p_{k}^2}{2m_1}
\nonumber \\ &&
+\Phi(r_{1},\dots, r_{N_1},r_{1}^\prime,\dots, r_{N_2}^\prime r_{1}^{\prime\prime},\dots, r_{N_3}^{\prime\prime})\biggr] T^{-1}\biggr),
\end{eqnarray}
where $N!$ takes the distinguish-ability
of classical galaxies into account, and $\Lambda$ is the normalization factor which results from integration over momentum space.
Now, integrating the momentum space, we obtain the following expression
\begin{eqnarray}
Z_{N}(T,V)&=& \frac{1}{\Lambda^{3N}N!}(2\pi m_{1}T)^{3N_{1}/2}(2\pi m_{2}T)^{3N_{2}/2}(2\pi m_{3}T)^{3N_{3}/2}
\nonumber \\ &&
\times Q_{N}(N_1,N_2,N_3,T,V),
\end{eqnarray}
where $Q_{N}(N_1,N_2,T,V)$ is the configurational integral given by \cite{ahm02},
\begin{eqnarray}
Q_{N}(N_1,N_2,N_3,T,V)&=&\int exp\biggl(-\Phi(r_{1},\dots, r_{N_1},r_{1}^\prime,\dots, r_{N_2}^\prime,r_{1}^{\prime\prime},\dots, r_{N_3}^{\prime\prime})\biggr)
\nonumber \\ &&
d^{3N_1}rd^{3N_2}r^{\prime}d^{3N_3}r^{\prime\prime}
\end{eqnarray}
where, after taking the softening parameter into account, we have
\begin{eqnarray}
\phi_{i,j}=-\frac{Gm^2}{(r_{ij}^2+\epsilon^2)^{1/2}}\biggl(1+\alpha-\alpha e^{-\mu{r_{ij}}}\biggr).
\end{eqnarray}
We can use  a two particle Mayer function
$f_{ij}=e^{-\Phi_{ij}/T}-1$, such that it
vanishes  in absence of interactions,  and  is non-zero only for interacting galaxies and the configurational integral
can be expressed in terms of the MOG modified function $f_{ij}$
\begin{eqnarray}
Q_{N}(T,V)&=& \int....\int \biggl((1+f_{12})(1+f_{13})\dots)\biggr)\nonumber\\
&\times&d^{3}r_{1}\dots d^{3}r_{N_1}d^{3}r_{1}^\prime \dots d^{3}r_{N_2}^\prime d^{3}r_{1}^{\prime\prime} \dots d^{3}r_{N_3}^{\prime\prime}
\end{eqnarray}
where we have used the MOG modified potential
\begin{equation}
f_{ij}=\frac{Gm_{1}m^\prime}{(r^2+\epsilon^2)^{1/2}}\big(1+\alpha-\alpha e^{-\mu r}\big).
\end{equation}
It may be noted that this configurational integral has been studied for usual Newtonian gravity \cite{ahm02}, however, here we have analyzed
the large distance  modification to it from MOG.
Here  $m^\prime$ can be $m_{1}$ or $m_{2}$, and so we can obtain,
\begin{eqnarray}
Q_{3}(T,V)&=& 4\pi V\int_{0}^{R_{1}}\biggl(1+\frac{Gm_{1}m_{2}}{T}\big[(1+\alpha)\frac{r^{2}}{(r^{2}+\epsilon^{2})^{1/2}}-\alpha\frac{r^{2}e^{-\mu r}}{(r^{2}+\epsilon^{2})^{1/2}}\bigr]\biggr)
\nonumber\\
 &\times&\biggl(1+\frac{Gm_{1}m_{3}}{T}\big[(1+\alpha)\frac{r^{2}}{(r^{2}+\epsilon^{2})^{1/2}}-\alpha\frac{r^{2}e^{-\mu r}}{(r^{2}+\epsilon^{2})^{1/2}}\bigr]\biggr)drdr^{\prime}dr^{\prime\prime}.\nonumber\\
\end{eqnarray}
In order to calculate the integral containing exponential part, we further make an approximation that for small
$\epsilon$, we have $(r^2+\epsilon^2)^{1/2}=r$, and obtain
\begin{eqnarray}
Q_{3}(T,V)=V^3\big(1+(\alpha_{1}+\beta_{1})\frac{m_2}{m_1}x\big)\big(1+(\alpha_{1}+\beta_{1})\frac{m_3}{m_1}x\big),
\end{eqnarray}
where $x=\beta\rho T^{-3}$ with $\beta=(3Gm_{1}^{2}/2)$.
The values of $\alpha_1$ and $\beta_1$ are given by
\begin{eqnarray}
\alpha_1&=&(1+\alpha)\gamma_{1},
\\
\beta_{1}&=&-\alpha\gamma_{2},
\end{eqnarray}
where,  we have
\begin{eqnarray}
\gamma_{1}&=&\biggl(\sqrt{1+\epsilon^{2}/R_{1}^{2}}+\frac{\epsilon^2}{R_{1}^2}ln{\frac{\epsilon/R_{1}}{1+\sqrt{1+\frac{\epsilon^2}{R_{1}^{2}}}}}\biggr),
\\
\gamma_{2}&=&\frac{2}{R_{1}^2\mu^2}\biggl(-e^{-\mu R_{1}}(\mu R_{1}+1)+1)\biggr).
\end{eqnarray}
By using the scale invariance,
 $\rho\to\lambda^{-3}\rho $,
$T\to\lambda^{-1}T$ and $r\to\lambda r$, we  obtain
\begin{equation}
\frac{3}{2}(Gm^2)^{3}\rho T^{-3}=\beta\rho T^{-3}.
\end{equation}
The general configurational integral for a three component system can be obtained by similar procedure,
thus we write expression for general term as
\begin{equation}
Q_{N}(T,V)=V^N\big(1+(\alpha_{1}+\beta_{1})x\big)^{N_1-1}\big(1+(\alpha_{1}+\beta_{1})\frac{m_2}{m_1}x\big)^{N_2}\big(1+(\alpha_{1}+\beta_{1})\frac{m_3}{m_1}x\big)^{N_3}.
\end{equation}
For $N_{1}>>1$, we can write the configurational integral as,
\begin{equation}
Q_{N}(T,V)=V^N\big(1+(\alpha_{1}+\beta_{1})x\big)^{N_1}\big(1+(\alpha_{1}+\beta_{1})\frac{m_2}{m_1}x\big)^{N_2}\big(1+(\alpha_{1}+\beta_{1})\frac{m_3}{m_1}x\big)^{N_3}.
\end{equation}
For a multi-component system i.e.,  the system containing $N_{1}$ galaxies with mass $m_{1}$, $N_{2}$
galaxies with mass $m_{2}$, $N_{3}$ galaxies with mass $m_{3}$,...,  the configurational integral can be generalized as,
\begin{equation}
Q_{N}(T,V)=V^N\prod_{l=1}^{s}\big(1+(\alpha_{1}+\beta_{1})\frac{m_l}{m_1}x\big)^{N_l}.
\end{equation}
Hence, the gravitational partition function for such a multi-component system given by,
\begin{equation}
Z_{N}(T,V)=\frac{1}{N!}\prod_{l}{N_{l}!Z_{N_{l}}},
\end{equation}
where we can write
\begin{equation}
Z_{N_{l}}=\frac{V^{N_{l}}}{\Lambda^{3N_{l}}N_{l}!}(2\pi m_{l})^{3N_{l}/2}\bigl(1+\frac{m_l}{m_1}(\alpha_1+\beta_1)x\bigr)^{N_{l}}.
\end{equation}
We will use partition function to study thermodynamics of the system.
\section{Multi-Component System of Galaxies}
\begin{figure}[h!]
 \begin{center}$
 \begin{array}{cccc}
\includegraphics[width=50 mm]{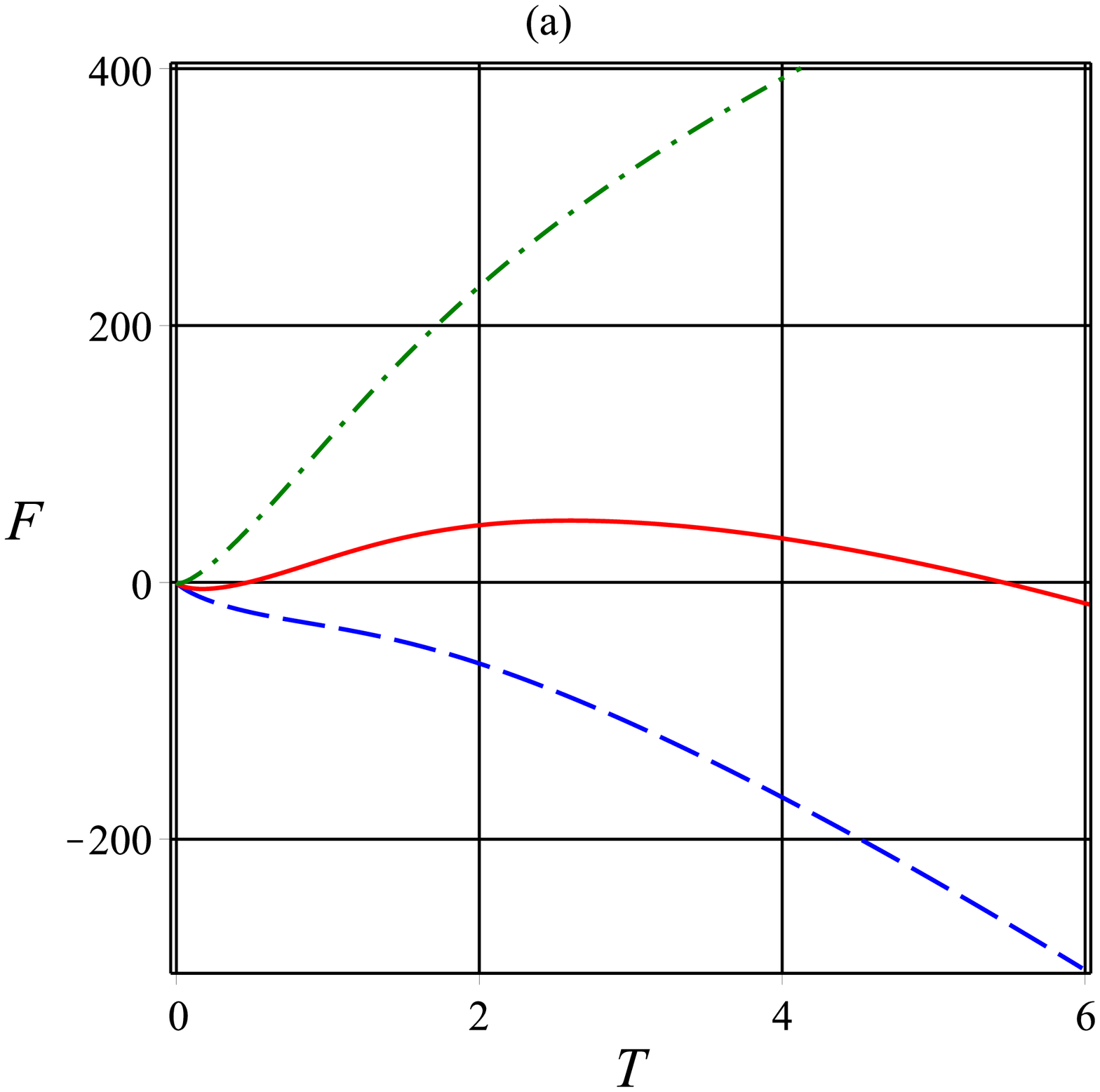}&\includegraphics[width=50 mm]{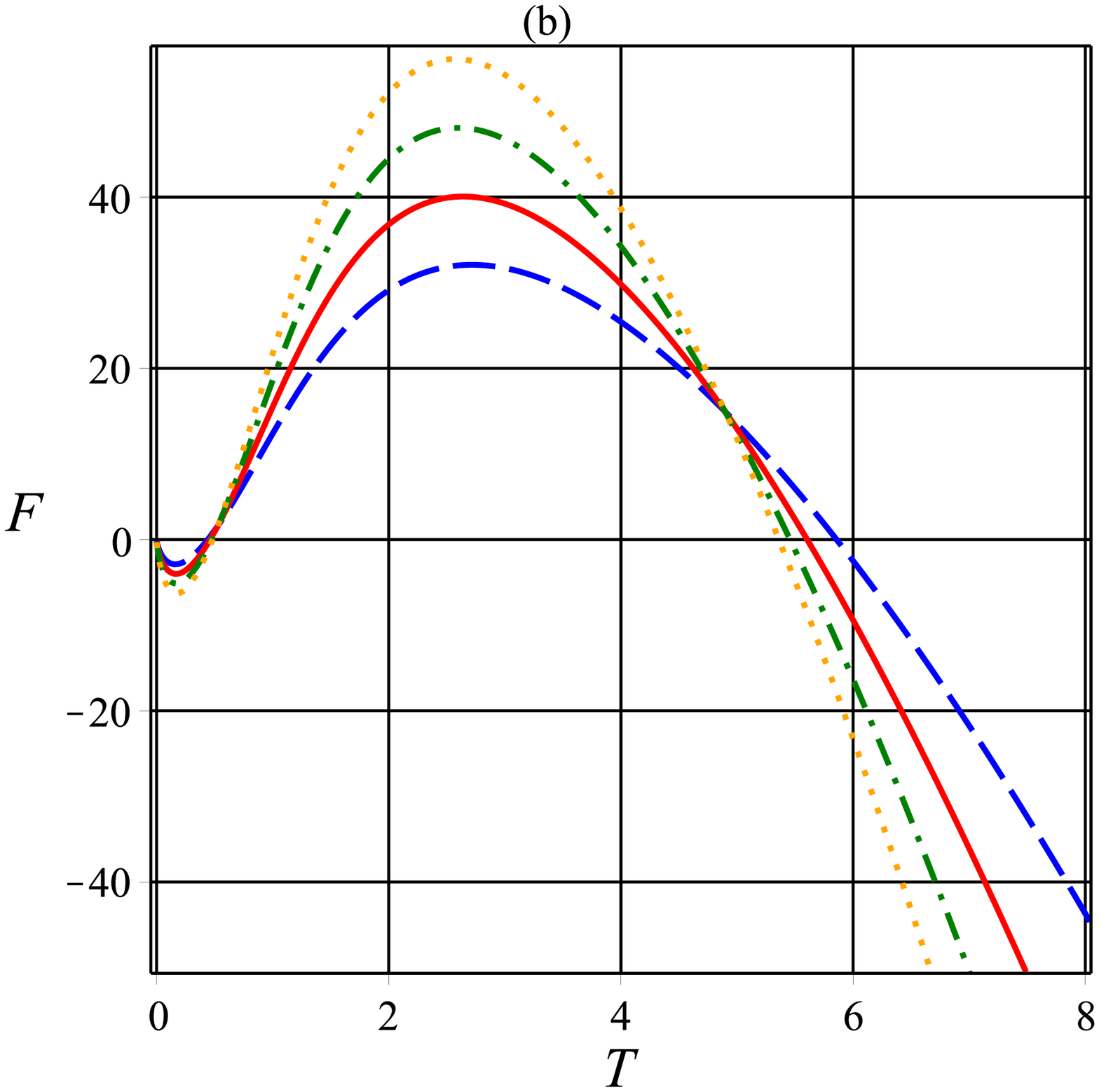}
 \end{array}$
 \end{center}
\caption{Typical behavior of the Helmholtz free energy in terms of $T$. (a) $N_{l}=3$ (blue dash),
$N_{l}=4$ (red solid), $N_{l}=5$ (green dash dot); $l=1, 2, 3, 4, 5$. (b) $N_{l}=4$; $l=1, 2, 3$ (blue dash),
$l=1, 2, 3, 4$ (red solid), $l=1, 2, 3, 4, 5$ (green dash dot); $l=1, 2, 3, 4, 5, 6$ (orange
 dot).}
 \label{fig1}
\end{figure}

It is possible to study clustering of different types of galaxies using the partition function of a
multi-component  system, as different types of galaxies can be modeled using a multi-component system \cite{ahm14, mu12, mu14}.
So, such a partition function can now be used to calculate relevant thermodynamical quantities for the
multi-component system of galaxies interacting with MOG potential. First of all, we can  write the
Helmholtz free energy using the following relation in the canonical ensemble,
\begin{eqnarray}
F&=&-T\ln Z_{N}(T,V)
 \nonumber \\ &=&-T\ln{\frac{1}{N!}\prod_{l}{N_{l}!Z_{N_{l}}}}
\nonumber  \\&=& NT\ln N-NT-T\sum_l \ln N_l+\sum_{l}{F_{l}}.
\end{eqnarray}
where
\begin{eqnarray}
\sum_l F_{l}=&-&\sum_{l}T\ln Z_{N_l}(T,V)\nonumber \\
&&-\sum_l \biggl(T\ln \biggl(\frac{1}{N_{l}!}\big(\frac{2\pi m_lT}{\Lambda^2}\big)^{3N_{l}/2}V^{N_{l}}\big(1+(\alpha_{1}+\beta_{1})\frac{m_l}{m_1}x\big)^{N_l}\biggr)\biggr).\nonumber\\
\end{eqnarray}
Here, we have  made use of Stirling's approximation,
$
\ln N!\approx N\ln N -N.
$
We can see behavior of the Helmholtz free energy in terms of temperature by Fig.\ref{fig1}.
We assumed different values for $N_{l}$, and find form the Fig. \ref{fig1} (a) that its value is
important in behavior of $F$. For $N_{l} = 3$, the Helmholtz free energy is completely negative.
For $N_{l}=4$, we can see some positive values of $F$, including a maximum, also a minimum for low temperature
case (Fig. \ref{fig1} (b)). In the case of high temperature, we can find large value for the Helmholtz
free energy. The value of the mentioned maximum of the Helmholtz free energy  depends on number of components.
Increasing number of components, increased value of the Helmholtz free energy at peak. There are some
critical temperatures ($T\approx0.5$ and $T\approx5$ in the Fig. \ref{fig1} (b)), where the Helmholtz
free energy of all multi-component systems are the same. Moreover, the Helmholtz free energy is zero at zero-temperature limit.

\begin{figure}[h!]
 \begin{center}$
 \begin{array}{cccc}
\includegraphics[width=60 mm]{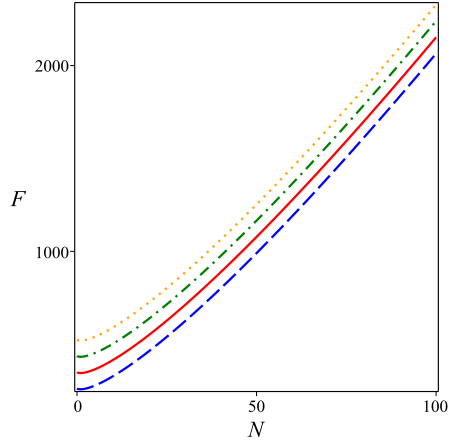}
 \end{array}$
 \end{center}
\caption{Typical behavior of the Helmholtz free energy in terms of $N$.  $N_{l}=5$; $l=1, 2, 3$ (blue dash), $l=1, 2, 3, 4$
(red solid), $l=1, 2, 3, 4, 5$ (green dash dot); $l=1, 2, 3, 4, 5, 6$ (orange
 dot).}
 \label{fig2}
\end{figure}

Also, the  behavior of the Helmholtz free energy in terms of $N$ is shown in Fig. \ref{fig2}.
It is clear that the Helmholtz free energy is an increasing function of $N$. As before, we can see that by increasing number of components,
value of the Helmholtz free energy is increased. We also find that the Helmholtz free energy is a decreasing function of $\alpha$.\\

\begin{figure}[h!]
 \begin{center}$
 \begin{array}{cccc}
\includegraphics[width=35 mm]{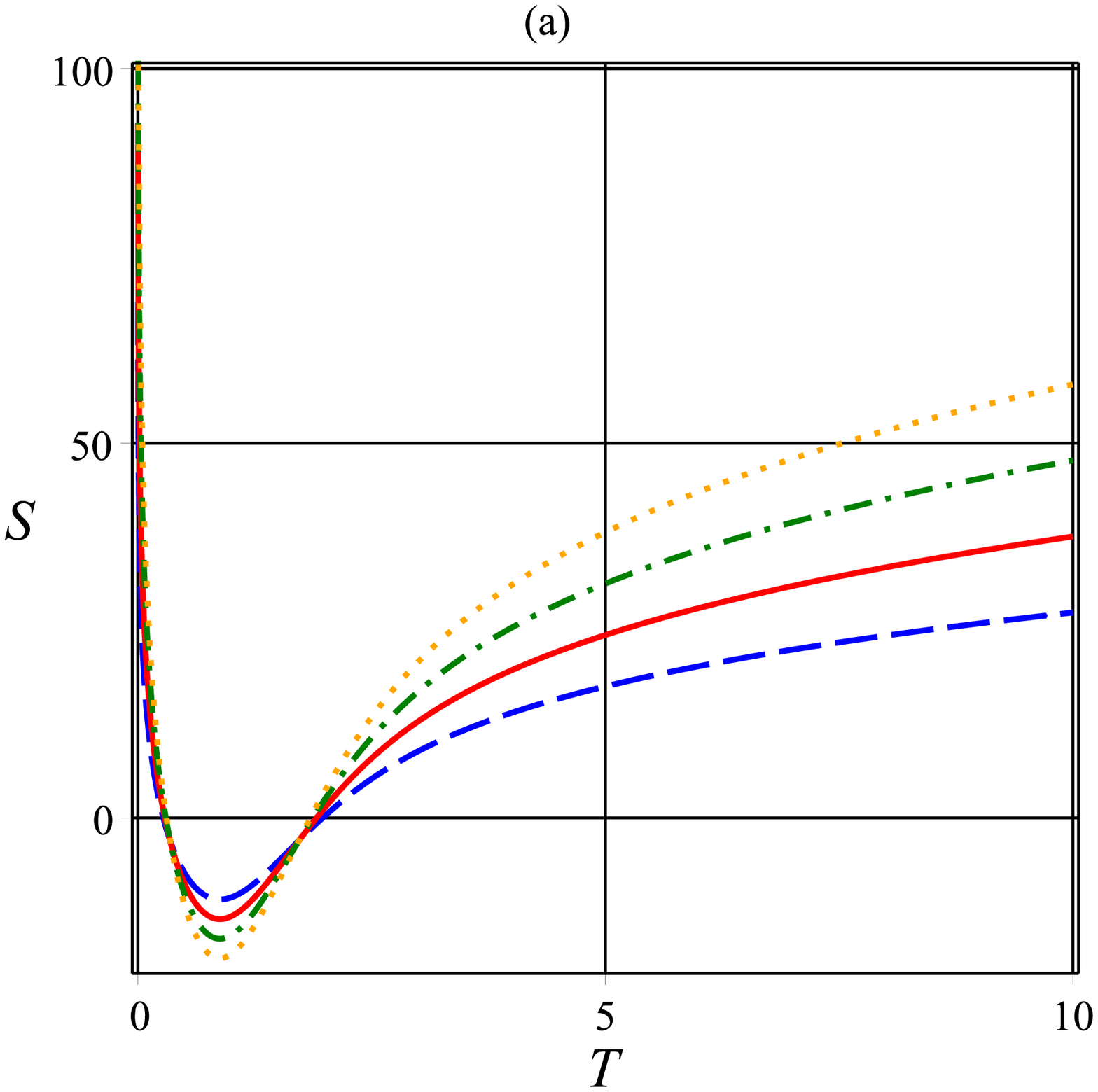}&\includegraphics[width=35 mm]{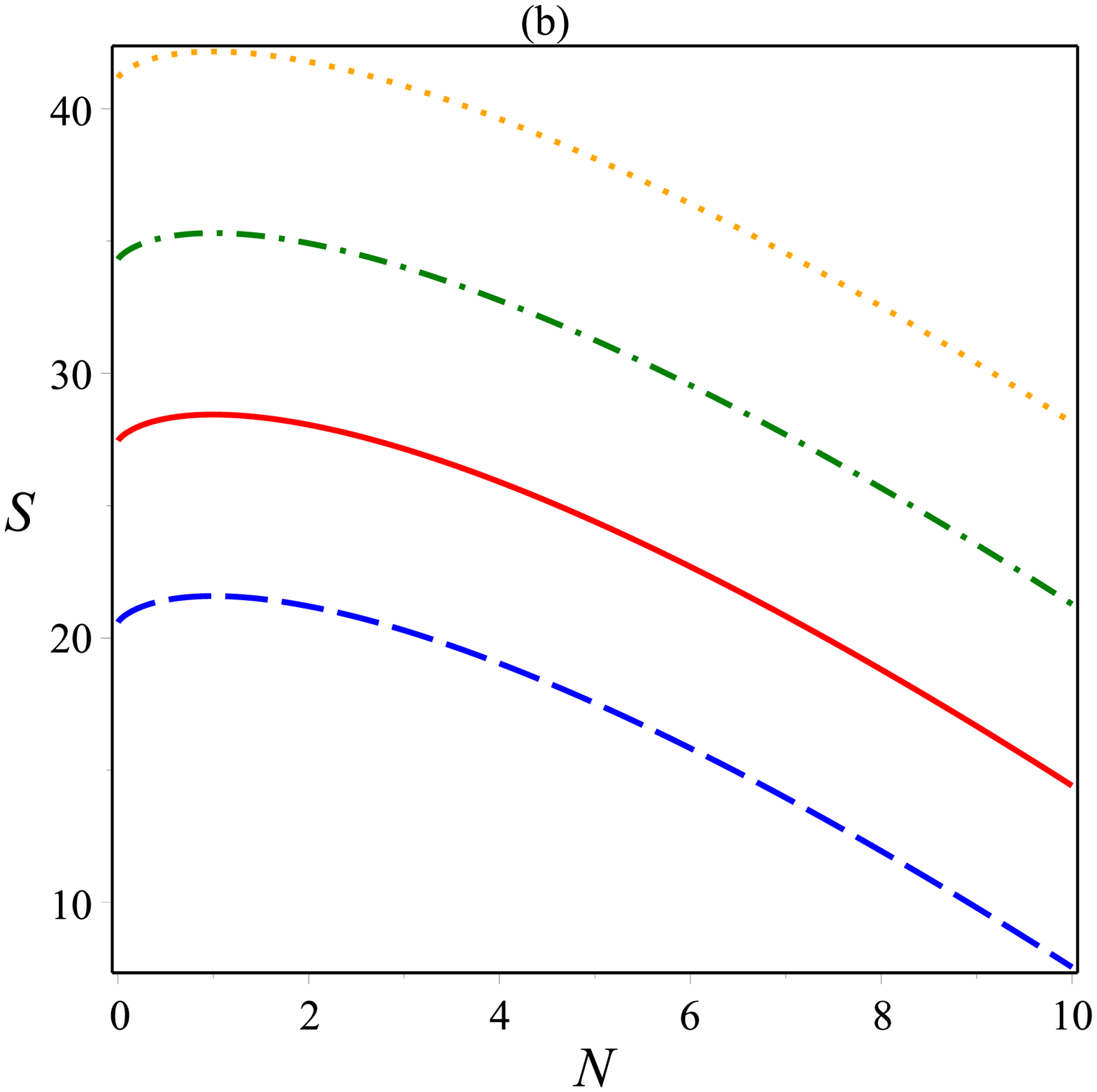}&\includegraphics[width=35 mm]{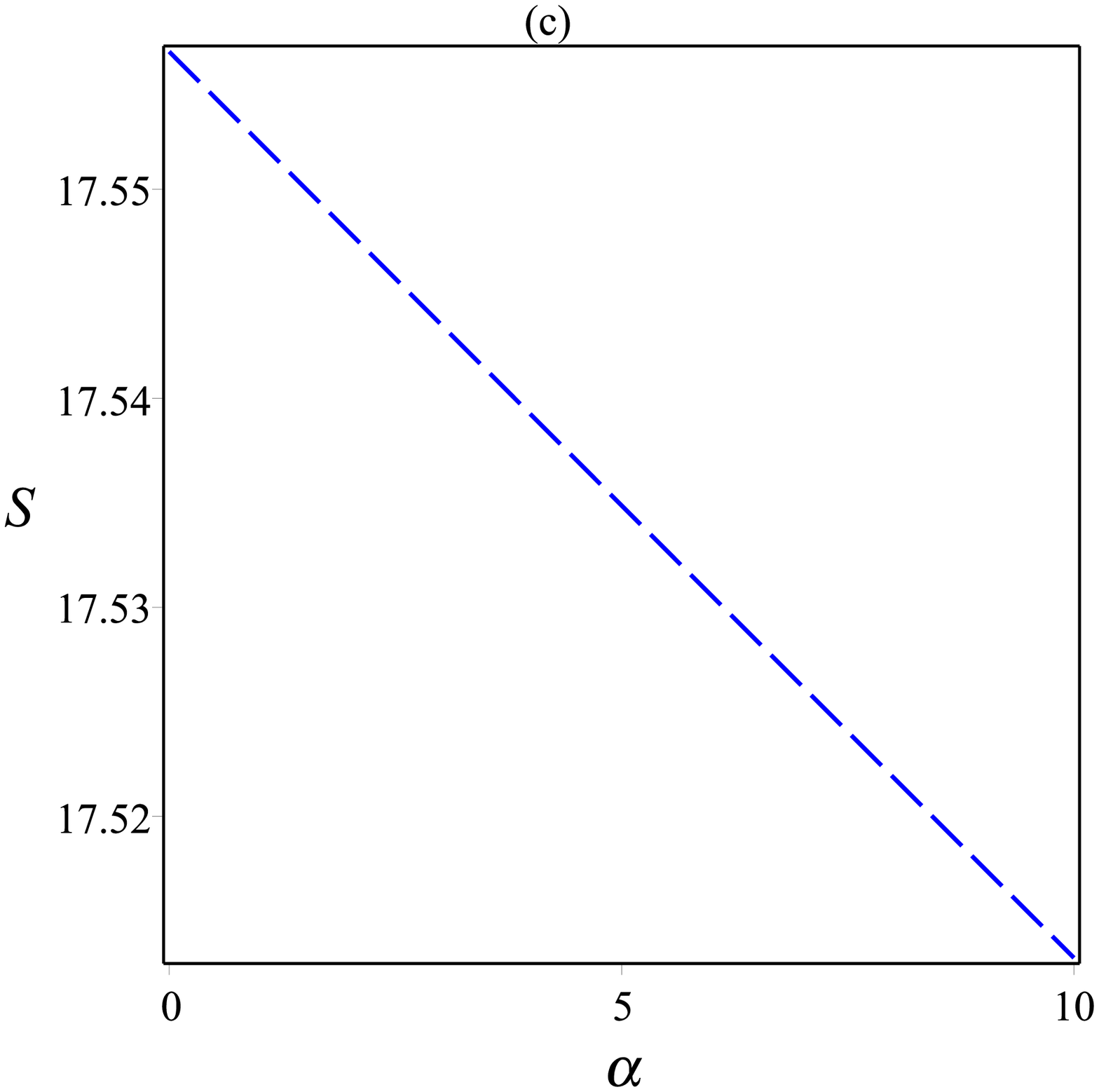}
 \end{array}$
 \end{center}
\caption{Typical behavior of the entropy in terms of (a) $T$, (b) $N$ and (c) $\alpha$.  $N_{l}=3$; $l=1, 2, 3$
(blue dash), $l=1, 2, 3, 4$ (red solid), $l=1, 2, 3, 4, 5$ (green dash dot); $l=1, 2, 3, 4, 5, 6$ (orange
 dot).}
 \label{fig3}
\end{figure}

The entropy $S$ can now be calculated from the Helmholtz free energy,
\begin{eqnarray}
S&=&-\biggl(\frac{\partial F}{\partial T}\biggr)_{N,V}
 \nonumber \\
&=&-N\ln N+N+\sum_l\ln N_l+\sum_{l}S_l,
\end{eqnarray}
where, we have
\begin{eqnarray}
\sum_l S_l&=& \sum_l \biggl(N_l\ln (\frac{T^{3/2}}{\rho})+N_{l}\ln \big(1+(\alpha_{1}+\beta_{1})\frac{m_l}{m_1}x\big) \nonumber \\ &&
-3N_l\frac{(\alpha_{1}+\beta_{1})\frac{m_l}{m_1}x}{1+(\alpha_{1}+\beta_{1})\frac{m_l}{m_1}x}
 +\frac{5}{2}N_l+\frac{3}{2}N_l\ln \big(\frac{2\pi m_l}{\Lambda^2}\big)\biggr).
\end{eqnarray}
In Fig. \ref{fig3}, we see typical behavior of the entropy. Fig. \ref{fig3} (a) shows that the entropy may be
negative at low temperature physics. It may be cause of some instability below a critical temperature. Also,
at the critical temperature, all multi-component systems are the same. Fig. \ref{fig3} (b) shows that value of
the entropy increases by increasing number of components. Finally,
Fig. \ref{fig3} (c) shows that the entropy is decreasing function of $\alpha$, by the small variation linearly.

\begin{figure}[h!]
 \begin{center}$
 \begin{array}{cccc}
\includegraphics[width=60 mm]{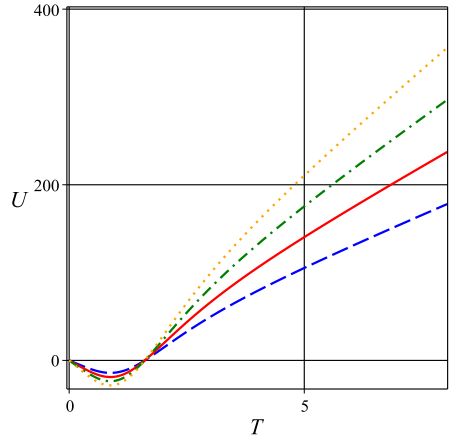}
 \end{array}$
 \end{center}
\caption{Typical behavior of the internal energy in terms of $T$.  $N_{l}=5$; $l=1, 2, 3$ (blue dash),
$l=1, 2, 3, 4$ (red solid), $l=1, 2, 3, 4, 5$ (green dash dot); $l=1, 2, 3, 4, 5, 6$ (orange
 dot).}
 \label{fig4}
\end{figure}

Here, we define the multi-component clustering parameter as,
\begin{eqnarray}
B_l=\sum_l\frac{(\alpha_{1}+\beta_{1})\frac{m_l}{m_1}x}{1+(\alpha_{1}+\beta_{1})\frac{m_l}{m_1}x},
\end{eqnarray}
We see the clustering parameter depends upon the masses of the interacting galaxies. This can be used to study the merging of galaxies.
The internal energy $U = F+TS$ of a multi-component system of galaxies, can now be expressed as,
\begin{equation}
U = \sum_l{\frac{3}{2}N_lT\big(1-2B_l\big)},
\end{equation}
which is independent of $N$, and decreasing function of $\alpha$. In the Fig. \ref{fig4}, we can see typical behavior
of the internal energy with respect to the temperature. We can see a minimum of energy at low temperature, and large
energy at high temperature. However, such minimum has negative internal energy, and negative entropy
(Fig. \ref{fig3} (b)). Hence, we can see negative entropy and internal energy below a critical temperature
($T_{c}\approx1$ with fixed parameters as given by figures), which may be sign of thermodynamical instability.

\begin{figure}[h!]
 \begin{center}$
 \begin{array}{cccc}
\includegraphics[width=45 mm]{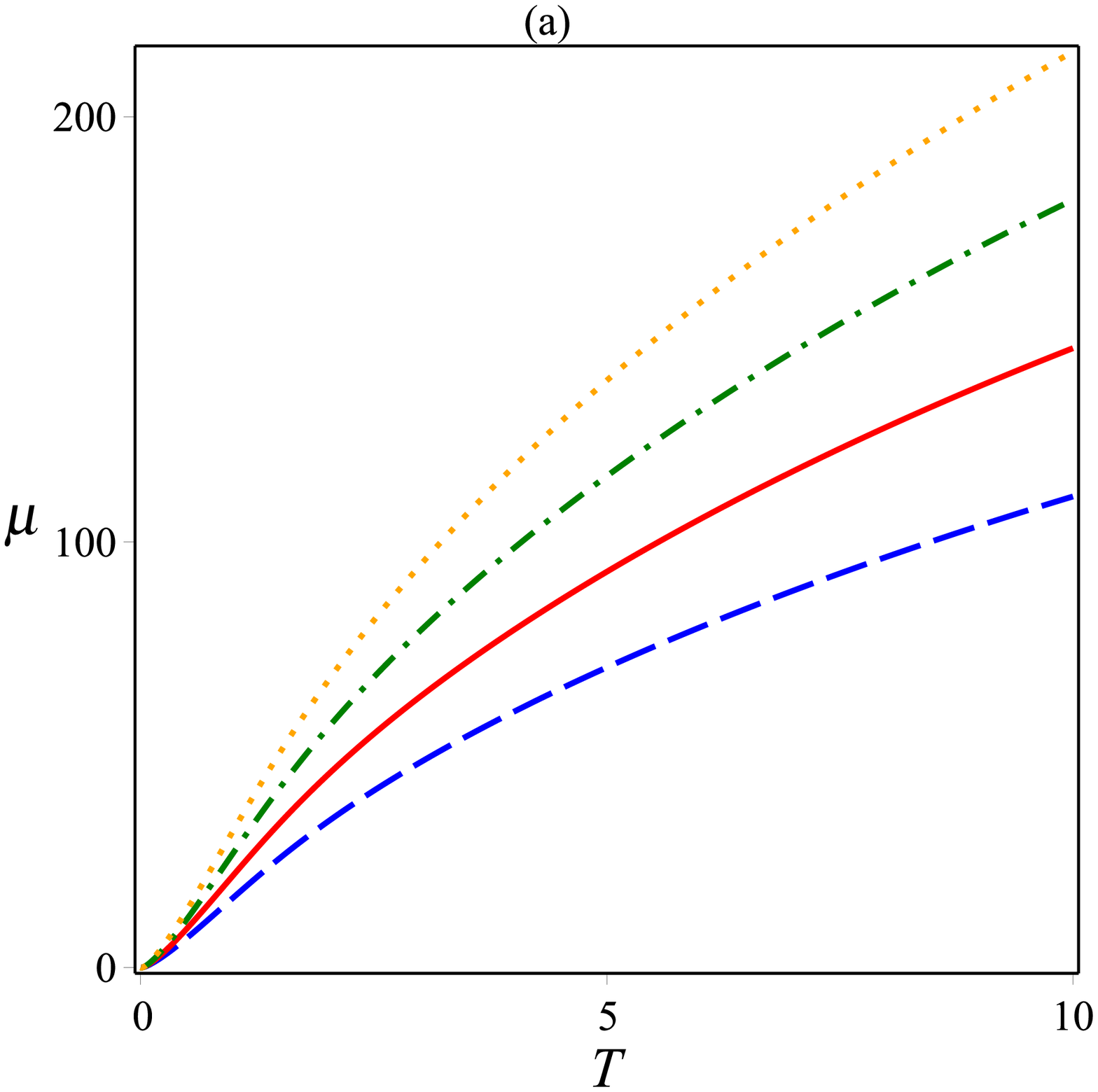}&\includegraphics[width=45 mm]{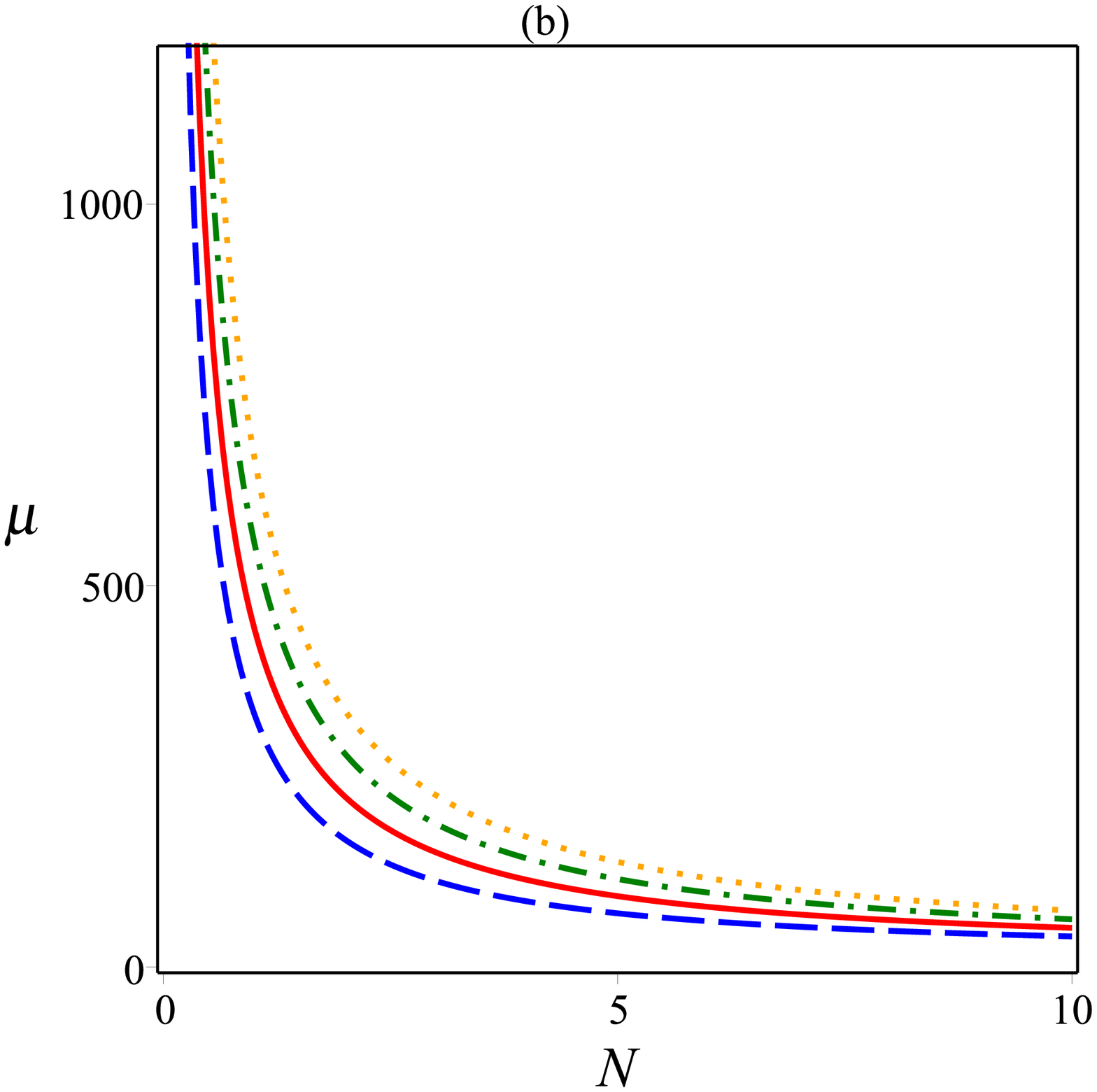}\\
\includegraphics[width=45 mm]{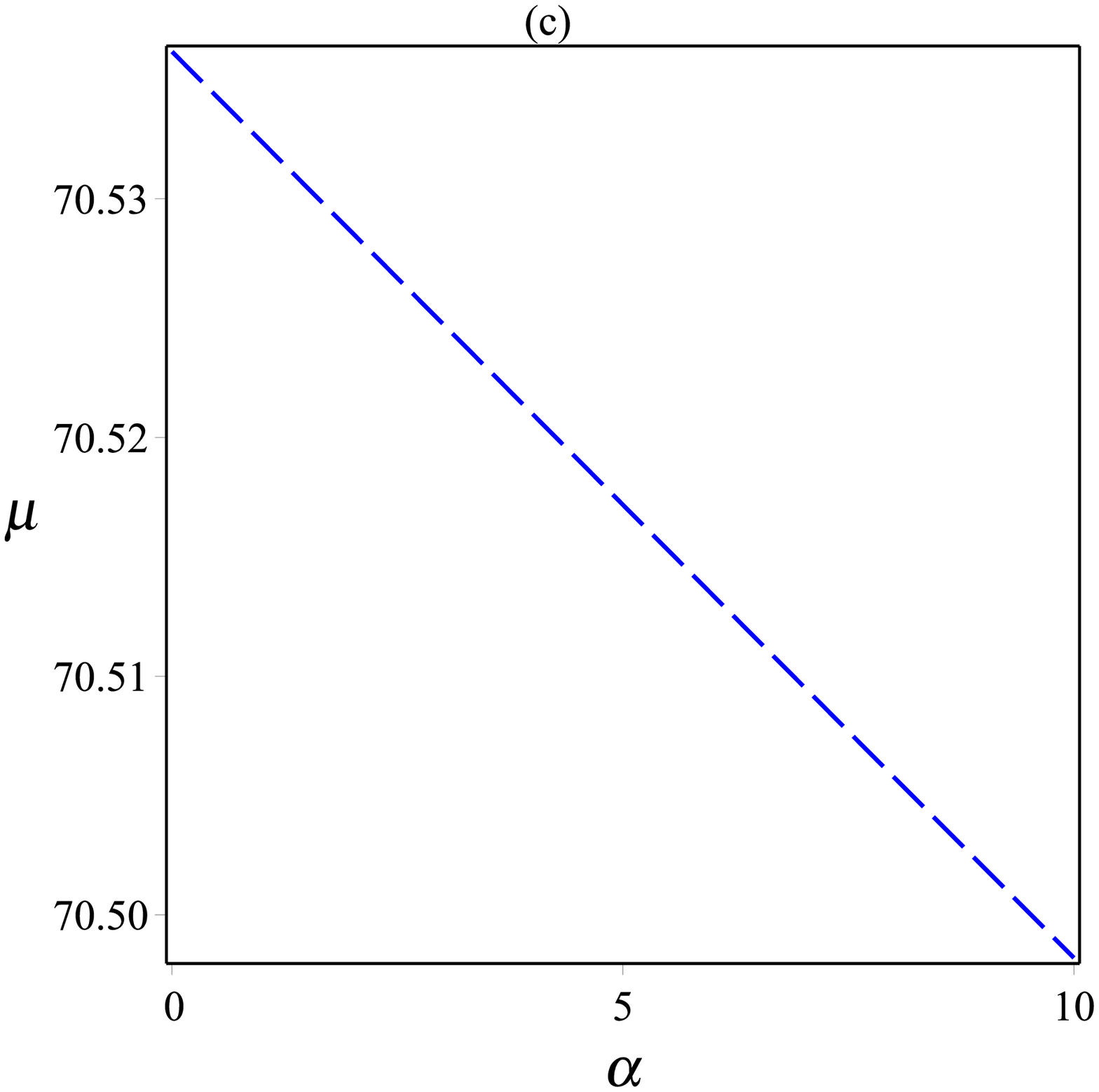}&\includegraphics[width=45 mm]{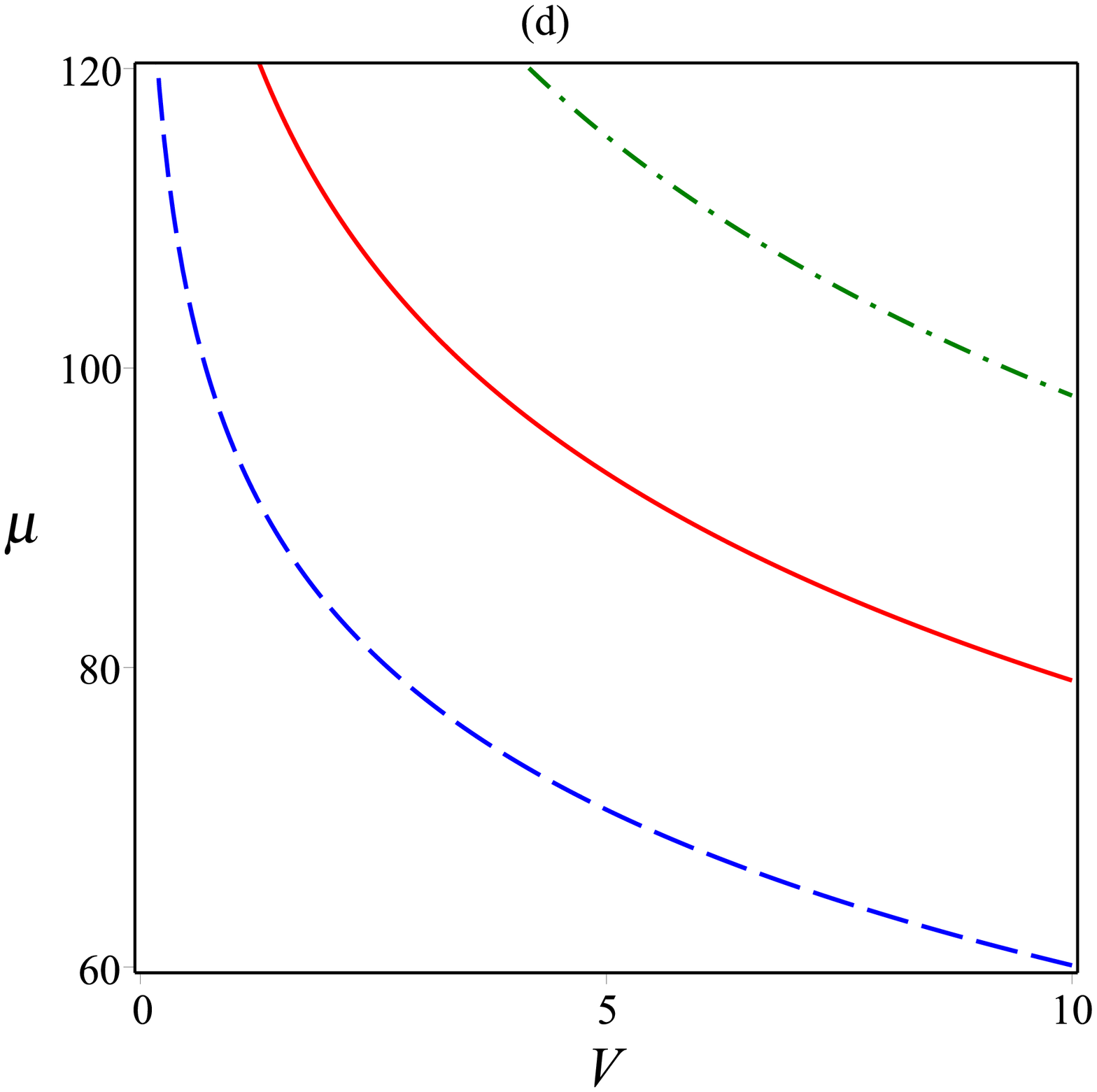}
 \end{array}$
 \end{center}
\caption{Typical behavior of the chemical potential in terms of (a) $T$, (b) $N$, (c) $\alpha$ and (d) $V$.
$N_{l}=5$; $l=1, 2, 3$ (blue dash), $l=1, 2, 3, 4$ (red solid), $l=1, 2, 3, 4, 5$ (green dash dot); $l=1, 2, 3, 4, 5, 6$ (orange
 dot).}
 \label{fig5}
\end{figure}

Similarly, we can write the  pressure $P$ and chemical potential $\mu$ as follows,
\begin{eqnarray}
P&=& -\biggl(\frac{\partial F}{\partial V}\biggr)_{N,T}\nonumber \\
&=& \sum_l\frac{N_lT}{V}\big(1-B_l\big),
 \\
\frac{N\mu}{T}&=&\biggl(\frac{F}{T}+\frac{PV}{T}\biggr)\nonumber\\
&=& \ln\frac{N!}{\prod N_{l}!}+\sum_l\frac{N_l\mu}{T},
 \\
\mu&=& \frac{T}{N}\left[N\ln N-N-\sum_l\ln N_l+\sum_l\frac{N_l\mu_{l}}{T}\right],
\end{eqnarray}
where
\begin{eqnarray}
\sum_l\frac{N_l\mu_l}{T}&=&\sum_l\biggl( N_l \ln (\rho T^{-3/2})- N_l \ln \big(1+(\alpha_{1}+\beta_{1})\frac{m_l}{m_1}x\big)
 \nonumber \\  &&
 - N_l \frac{3}{2}\ln \big(\frac{2\pi m_l}{\Lambda^2}\big)- N_l B_l\biggr).
\end{eqnarray}
Now Fig. \ref{fig5} show typical behavior of the chemical potential in terms of $T$, $N$, $\alpha$ and $V$. In the Fig.
\ref{fig5} (a) we can see that chemical potential is increasing function of the temperature. Also, increasing number
of component increases value of the chemical potential. From the Fig. \ref{fig5} (b) we can see that chemical potential
is decreasing function of $N$. Fig. \ref{fig5} (c) shows that chemical potential is linearly decreasing function of $N$.
Finally, Fig. \ref{fig5} (d) shows that chemical potential is decreasing function of $V$.

The probability of finding $N$ galaxies can be written as
\begin{eqnarray}
F(N)=\frac{e^{\frac{N\mu}{T}}Z_{N}(V,T)}{Z_{G}(T,V,z)},
\end{eqnarray}
where $Z_{G}=zZ_{N}$ and $z$ is the activity.\\
Thus, for a multi-component system of gravitationally interacting galaxies of different species,  we have,
\begin{eqnarray}
F(N_1,N_2,\dots N_l)&=& \prod_l \frac{Z_{N_l}e^{{\frac{N_l \mu}{T}}}}{e^{\bar N_l(1-B_l)}}
\end{eqnarray}
where $N=N_1+N_2+\dots N_l$.
The general distribution of a multi-component system with MOG effect can be written as,
\begin{eqnarray}
F(N) &=& \frac{\bar{N_1}^{N_1}}{N_1!}\biggl(1+\frac{N_1}{\bar N_1}\frac{B_1}{(1-B_1)}\biggr)^{N_{1}-1}\biggl(1+\frac{B_1}{(1-B_1)}\biggr)^{-N_1}
\nonumber \\  && \times e^{(-N_1B_1-\bar {N_1}(1-B_1))}
 \nonumber \\ && \times \prod_{2}^l\biggl(\frac{\bar{N_l}^{N_l}}{N_l!}\biggl(1+\frac{N_l}{\bar N_l}\frac{B_l}{(1-B_l)}\biggr)^{N_{l}}\biggl(1+\frac{B_l}{(1-B_l)}\biggr)^{-N_l}
\nonumber \\  && \times e^{(-N_lB_l-\bar {N_l}(1-B_l))}\biggr)
\end{eqnarray}
If all the galaxies are of same mass the result reduces to
\begin{eqnarray}
F(N) &=& \frac{\bar{N}^{N}}{N!}\biggl(1+\frac{N}{\bar N}\frac{B}{(1-B)}\biggr)^{N-1}\biggl(1+\frac{B}{(1-B)}\biggr)^{-N}
\nonumber \\  && \times e^{(-NB-\bar N(1-B))},
 \nonumber \\
\end{eqnarray}
where, we have
\begin{equation}
B=\frac{(\alpha_{1}+\beta_{1})x}{1+(\alpha_{1}+\beta_{1})x}.
\end{equation}
We find that distribution function is increasing function of temperature,  while it  is a decreasing function of numbers.
We also find that clustering parameter decreases value of the distribution function.
In the next section the behavior of the above mentioned parameter for the multi-component system is discussed.

We can  defined the clustering parameter between the galaxies of different mass components as follow,
\begin{equation}
B_l=\sum_l\frac{(\alpha_{1}+\beta_{1})\frac{m_l}{m_1}x}{1+(\alpha_{1}+\beta_{1})\frac{m_l}{m_1}x},
\end{equation}
We see the clustering parameter depends upon the masses of the interacting galaxies. This can be used to study the merging of galaxies.
We can express the parameter $x$,  in terms of the number of galaxies as
\begin{equation}
x=\beta\rho T^{-3}=\beta (N_l/N_1)\bar{\rho}T^{-3}=(N_l/N)y.
\end{equation}
Thus, we can write
\begin{eqnarray}
B_l=\sum_l\frac{(\alpha_{1}+\beta_{1})\frac{m_l N_l}{m_1 N_1}y}{1+(\alpha_{1}+\beta_{1})\frac{m_l N_l}{m_1 N_1}y},
\end{eqnarray}
It may be noted that as,  we  can express the clustering parameter as
\begin{equation}
B_l=b\sum_l\biggl(\frac{1+\alpha(1-\frac{\gamma_{2}}{\gamma_{1}})\frac{m_lN_l}{m_lN_l}}{1+
\alpha(1-\frac{\gamma_{2}}{\gamma_{1}})\frac{m_lN_l}{m_lN_l}b}\biggr),
\end{equation}
So, for  one component,  we have
\begin{equation}
B=b\frac{1+\alpha(1-\frac{\gamma_{2}}{\gamma_{1}})}{1+\alpha(1-\frac{\gamma_{2}}{\gamma_{1}})b}.
\end{equation}
where, $b$ is given by
\begin{equation}
b=\frac{\gamma_{1}y}{1+\gamma_{1}y},
\end{equation}
It is interesting to note that the effect of  MOG modified potential enter into the distribution function only through the clustering parameter.
Now for attractive potential, we have  $0\le B_l\le 1$.
So, it is the deviation of distribution function, which can give an estimate of merging as well as multi-component clustering.
Thus, it is possible to study the clustering of galaxies of different masses using a multi-component systems.

\section{Conclusion and Discussion}
In this paper, we have studied the clustering of a system of galaxies interacting thought a MOG modified Newtonian potential.
As it is possible for the system of galaxies to have different masses, we have analyzed this system using a
  multi-component systems.  This MOG modified Newtonian  potential can be obtained from the
  weak field approximation of MOG, and we have used it for calculating the    partition function
of this  multi-component system. We compute the  partition function, and studied the    thermodynamics of this system using that
partition function. We also analyzed the
 general clustering parameter for this multi-component system of galaxies interacting though MOG.

Indeed we have thermodynamical study of clustering of the multi-component systems of galaxies in modified gravity to see
how MOG (also number of components) affect thermodynamics quantities. We have shown that clustering parameter decreased
value of most important thermodynamics quantities, while number of components increase value of thermodynamics variables.
Helmholtz free energy for multi-component system of galaxies is evaluated and the variation Helmholtz free energy $F$
with temperature depends on the value of $N_l$. When $N_l = 3$, the $F$ is negative for all values of temperature.
For $N_l = 5$ the Helmholtz free has high positive values and keeps increasing for higher temperatures. For $N_l = 4$, $F$
has positive values with a minima and a maxima, we notice that changing the values of $l$ the maxima or peak shifts upwards
with increasing values of $l$. We also study the variation of free energy as a function of $N$, which increases with the value of $N$,
and by increasing the number of components the free energy curve shifts upwards. The entropy of multi-component system of galaxy is also studied,
it is seen that it has negative values for low values of temperature, and further increasing the temperatures the entropy becomes positive and
keeps increasing. The variation of entropy with $N$ shows that the value of entropy increases initially as a function of $N$, and then decreases
on further increasing the value of $N$. We also see that in entropy versus $N$ plot, increasing the value of $l$ from $3$ to $6$, and  the over all
entropy curve is shifted upwards. To check the dependency of entropy on the parameters $\alpha$, we plot $S$ as a function of $\alpha$ and
notice that $S$ decreases linearly with increasing value of $\alpha$. The study of the internal energy of multi-component system of galaxies
shows that it depends on the the multi-component clustering parameter, $B_l$. The behavior of internal energy with respect to temperature is
studied, and it is seen that at low temperature the internal energy has a minima. However,  as the temperature increases
further it increases and takes
large  values. Chemical potential plays very important role in clustering of galaxies, and it depends on the temperature, $\alpha$, $N$
and volume, $V$. The variation of chemical potential with respect to temperature shows that as temperature increases the chemical potential
increases and for higher values $l$,  the chemical potential increases rapidly. With $N$ the chemical potential initially drops rapidly  for
small values of $N$, and remains almost constant for higher values of $N$. Furthermore,  we notice that, the rate at which $\mu$ changes for
large values of $l$ is slower in comparison to small values of $l$. The chemical potential decreases linearly as the value of $\alpha$ increases.
The  chemical potential decreases logarithmically as the volume of the multi-component system increases. We
found that distribution function is increasing function of clustering parameter as well as temperature, while is decreasing function of numbers.

\end{document}